\documentclass[prd,11pt,twocolumn,nofootinbib,preprintnumbers]{revtex4}
\usepackage{amsmath,amssymb,slashed,mathtools,slashed}
\usepackage{graphicx,multirow}
\usepackage{units}
\usepackage[hyperfootnotes=false,colorlinks,citecolor=blue]{hyperref}
\usepackage{mathtools}
\usepackage{pifont}
\usepackage{amsmath,amssymb}
\usepackage{epsfig}
\usepackage{url}
\usepackage{subfigure}
\usepackage{graphicx}
\usepackage{grffile}
\usepackage[usenames,dvipsnames]{color}
\usepackage{slashed}
\usepackage{array}
\usepackage{multirow}
\usepackage{braket}
\usepackage[colorlinks,citecolor=blue]{hyperref}
\usepackage{color}
\usepackage{cancel}
\usepackage{gensymb}

\def\a {\alpha}
\def\b {\beta}

\def\l {\lambda}

\def\bar {\overline}

\def\be {\begin{equation}}
\def\ee {\end{equation}}
\def\beq {\begin{equation}}
\def\eeq {\end{equation}}
\def\bea {\begin{eqnarray}}
\def\eea {\end{eqnarray}}

\newcommand{\besub}{\begin{subequations}}
\newcommand{\eesub}{\end{subequations}}

\def\beq{\begin{equation}}
\def\eeq{\end{equation}}
\def\barr{\begin{array}}
\def\earr{\end{array}}

\begin{document}
\title{Single-step first order phase transition and gravitational waves in a SIMP dark matter scenario}

\author{Nabarun Chakrabarty}
\email{nabarunc@iitk.ac.in}
\affiliation{Department of Physics, Indian Institute of Technology Kanpur, Kanpur, Uttar Pradesh-208016, India} 

\author{Himadri Roy}
\email{himadri027roy@gmail.com}
\affiliation{Department of Physics, Indian Institute of Technology Kanpur, Kanpur, Uttar Pradesh-208016, India}

\author{Tripurari Srivastava}
\email{tripurarisri022@gmail.com}
\affiliation{Department of Physics and Astrophysics, University of Delhi, Delhi 110007, India}

\begin{abstract} 
We investigate the non-zero temperature dynamics of a sub-GeV dark matter scenario
freezing-out via self-interactions. As a prototype, we take up the case of a scalar dark matter species undergoing $3 \to 2$ number changing annihilations catalysed by another scalar. We study the shape of the thermal potential of this scenario in a parameter region accounting for the observed relic abundance. An analysis reveals the possibility of a first order phase transition with bubble nucleation occurring at sub-GeV temperatures. This finding can be correlated with the typical sub-GeV masses in the framework. The gravitational wave spectra associated with such a phase transition is subsequently computed. 

\end{abstract} 
\maketitle

\section{Introduction} 

It has been concurred by numerous observations~\cite{Rubin:1970zza, Bertone:2004pz, Hu:2001bc, WMAP:2012nax} that around a fourth of the total energy budget of the universe is some non-luminous matter known as dark matter (DM). The amount of such matter
in the universe is also well measured from various observations, e.g., cosmic microwave background (CMB)~\cite{Planck:2018vyg} and large scale structure surveys~\cite{Blumenthal:1984bp}.  If DM is thought to be an elementary particle, then there is no information available from the current experiments on its mass and quantum numbers. More importantly, that there is no such particle candidate for DM within the Standard Model (SM), advocates the presence of additional dynamics. While DM can be hypothesized to stem from some beyond-the-Standard Model (BSM) framework, a possibility in this direction is that DM is a weakly interacting massive particle (WIMP)~\cite{Kolb:1990vq, Jungman:1995df, Bertone:2004pz, Feng:2010gw, Arcadi:2017kky, Roszkowski:2017nbc}. That is, the DM-SM interactions are in the \emph{weak} ballpark. Moreover, the DM relic abundance in the WIMP paradigm is generated typically through $2_{\text{DM}} \to 2_{\text{SM}}$ annihilation processes followed by a thermal \emph{freeze out}. 

Search for DM through its scattering with nuclei at terrestrial detectors is known as direct detection~\cite{XENON:2018voc, LUX:2017ree}. However,
non-observation of such scattering at have put stringent upper limits on the corresponding cross sections, and ultimately, on the DM-SM interactions. Such small interactions are also consistent with the null results obtained in DM searches at the energy frontier, the Large Hadron Collider (LHC) being the prime example here~\cite{Abercrombie:2015wmb}. These results have pushed a plethora of WIMP scenarios to a corner~\cite{Roszkowski:2017nbc, Arcadi:2017kky}. An interesting alternative therefore could a strongly interacting massive particle (SIMP) \cite{Bernal:2015xba,Choi:2016hid,Choi:2016tkj,Bernal:2017mqb,Chauhan:2017eck,Mohanty:2019drv,Bhattacharya:2019mmy,Lee:2015gsa,Yamanaka:2015tba,Hochberg:2015vrg,Hochberg:2014kqa,Choi:2017zww,Choi:2018jsb,Herms:2018ajr,Ma:2015mjd}. In this case, the freeze-out dynamics is driven by DM matter \emph{number changing} processes, i.e., processes that feature an imbalance of DM particles in the initial and final states. The main advantage of the SIMP paradigm is that it allows such number changing processes to be predominantly driven by DM self interactions. The DM-SM interaction stengths are thus permitted to be appropriately small so as to conform with the limits from direct detection and colliders. A SIMP therefore typically has sub-GeV mass and a large self-scattering cross section. \vspace{-0.6mm}

With the advent of gravitational wave (GW) interferometry~\cite{Thrane:2013oya, Sathyaprakash:2012jk,LISA:2017pwj,Crowder:2005nr,Seto:2001qf,Sato:2017dkf, Lentati:2015qwp,Janssen:2014dka,Fairbairn:2019xog}, it has become possible to probe a dark sector through its GW imprints~\cite{Breitbach:2018ddu,Schwaller:2015tja, Jaeckel:2016jlh, Addazi:2016fbj,Baldes:2017rcu,Tsumura:2017knk,Baldes:2018emh,Croon:2018erz}. In this study, we aim to study the possibility of a strong first order phase transition (SFOPT)~\cite{Kamionkowski:1993fg,Witten:1984rs,Hogan:1986qda,Grojean:2006bp} and the GW spectrum from associated with a sub-GeV SIMP framework. The model we take up is where the SM is extended by the scalars 
$\phi$ and $\delta$ that are singlets under the SM gauge symmetry \cite{Mohanty:2019drv}. The scalar $\phi$, stabilised by a governing $\mathbb{Z}_2$ symmetry, becomes the DM candidate here. The number changing process for this case that is consistent with the $\mathbb{Z}_2$ is $\phi\phi\phi \to \phi h_2$, where $h_2$ is an admixture of $\delta$ and the Higgs doublet $H$. The scalar $\delta$ and its interaction with the DM $\phi$ thus play crucial roles in generating the observed relic density through the aforesaid $3 \to 2$ dynamics.
Given the importance of $\delta$ in this setup, a pertinent question to ask is whether the same can trigger a SFOPT and how strong is the resulting GW amplitude. We plan to explore this possibility here by incorporating finite temperature corrections to the scalar potential along the direction of 
$\delta$.

The study is organised as follows. We introduce the setup and outline the $3 \to 2$ dynamics in section \ref{Intro}. Thermal corrections to the scalar potential are computed in section \ref{sfopt}. The same section presents the GW amplitude for representative benchmark points. We finally conclude in section \ref{conclu}. Some important formulae can be found in the Appendix \ref{app}.

\section{The sub-GeV dark matter framework and $3 \to 2$ dynamics}\label{Intro}

We extend the SM with the real gauge singlet scalars $\phi$ and $\delta$ \cite{Mohanty:2019drv}. A $\mathbb{Z}_2$ symmetry is imposed under which $\phi \to -\phi$ while all other fields transform trivially. This negative $\mathbb{Z}_2$ charge for $\phi$ stabilises the same and makes it a potential DM candidate. The scalar potential of the setup reads
\bea
V & = & \frac{\mu_{\phi}^2}{2} \phi^2 + \frac{\mu_{\delta}^2}{2} \delta^2 + \mu_H^2 H^\dagger H + \frac{\mu_{21}}{2} \phi^2 \delta  + \frac{\mu_{22}}{3!} \delta^3 \nonumber \\
&& + \mu_{23}  H^\dagger H \delta 
+ \frac{\lambda_{11}}{4!} \phi^4 + \frac{\lambda_{12}}{4} \phi^2 \delta^2 + \frac{\lambda_{22}}{4!} \delta^4 \nonumber \\
&& + \frac{\lambda_{13}}{2} \phi^2 H^\dagger H +   \frac{\lambda_{23}}{2} \delta^2 H^\dagger H + \lambda_{33} \left[ H^\dagger H\right]^2. 
\eea
In the above, all parameteres are taken real and $H$ denotes the SM Higgs doublet. The doublet $H$ and the singlet $\delta$ receive vacuum expectation values (VEVs) $v$ and $v_\delta$ respectively. One then writes
\bea
H = \begin{pmatrix}
G^+ \\
\frac{1}{\sqrt2}(v + h + i G^0)
\end{pmatrix};~~~
\delta = v_\delta + \delta^\prime.
\eea 
Here, $v$ = 246 GeV. It is noted that $\phi$ does not receive a VEV and thus the 
$\mathbb{Z}_2$ remains intact. In all, such a configuration of the VEVs introduces a mixing between $h$ and $\delta^\prime$. The mixing of $\phi$ is however forbidden by the $\mathbb{Z}_2$ symmetry. The mass matrix in the $(h~~\delta^\prime)$ basis is
\bea
\mathcal{M}^{2} & = &\begin{pmatrix}
2 \lambda_{33} v^{2} &  \lambda_{23} \hspace{0.5mm} v v_{\delta}  + \mu_{23} v \\
\lambda_{23} \hspace{0.5mm} v v_{\delta} \hspace{0.5mm} + \mu_{23} v & \frac{ \mu_{22} v_{\delta}}{2} + \frac{\lambda_{22}}{3} v_{\delta}^{2} - \frac{\mu_{23}}{2} \frac{v^{2}}{v_{\delta}}
\end{pmatrix}.\label{mass_sq_mat}
\eea
We have made use of the tadpole conditions $\frac{\partial V}{\partial h} = 0 = \frac{\partial V}{\partial \delta^\prime}$
while deriving Eq.(\ref{mass_sq_mat}). One derives the tadpole conditions to be
\besub
\bea
\mu_{h}^{2} &= &- \big(\mu_{23} \hspace{1mm} v_{\delta} + \frac{\lambda_{23}}{2} \hspace{1mm} v_{\delta}^{2} + \lambda_{33} \hspace{1mm} v^{2}\big), \\ 
\mu_{\delta}^{2}  & = & - \big(\frac{\mu_{22}}{2} v_{\delta} + \frac{\lambda_{22}}{6} v_{\delta}^{2} + \frac{\mu_{23}}{2} \frac{v^{2}}{v_{\delta}} + \frac{\lambda_{23}}{2} v^{2} \big).
\eea
\eesub
Eq.(\ref{mass_sq_mat}) is diagonalised through a $2\times2$ rotation parameterised by a mixing angle $\theta$. That is,
\bea
\begin{pmatrix}
h \\ \delta^\prime
\end{pmatrix} = \begin{pmatrix}
\cos\theta & \sin\theta \\ 
-\sin\theta & \cos\theta
\end{pmatrix} \begin{pmatrix}
h_{1} \\ h_{2}
\end{pmatrix},
\eea
which leads to the eigenstates $(h_1,h_2)$. We identify $h_1$ with the discovered Higgs boson with $m_{h_1}$ = 125 GeV. The DM mass is given by  
\bea
m_\phi^2 = \mu^2_\phi + \mu_{21} v_\delta
+ \frac{1}{2}\l_{12}v^2_\delta
+ \frac{1}{2}\l_{13}v^2.
\eea

Some of the parameters in the scalar potential can be expressed in terms of the physical quantities of the theory such as masses and mixing angles using the relations
\besub
\bea
\lambda_{33} & = & \frac{(m_{h_{1}}^{2} \cos^{2}\theta + m_{h_{2}}^{2} \sin^{2}\theta )}{2\hspace{0.5mm} v^{2}}, \\
\mu_{23} & = & \frac{(m_{h_{2}}^{2} - m_{h_{1}}^{2}) \cos \theta \sin \theta }{v} - \lambda_{23}\hspace{0.5mm} v_{\delta}, \\
\mu_{22} & = &  \frac{2}{v_{\delta}}(m_{h_{1}}^{2} \sin^{2}\theta + m_{h_{2}}^{2} \cos^{2}\theta) - \frac{\lambda_{22}}{3}  \hspace{0.5mm} v_{\delta}\nonumber \\
&& + \mu_{23} \frac{v^{2}}{v^{2}_{\delta}}, \\
\mu_\phi^2 &=& m^2_\phi - \mu_{21} v_\delta
- \frac{1}{2}\l_{12}v^2_\delta
- \frac{1}{2}\l_{13}v^2.
\eea
\eesub
The independent variables of the present framework thus are
$\{m_{h_2}, m_\phi, \mu_{21}, v_h,
 v_\delta,\text{sin}\theta, \l_{11},\l_{12},\l_{13},\l_{22},\l_{23} \}$.

The relic abundance in a strongly interacting DM setup is generated through number changing processes. In this setup, the leading number changing processes consistent with the $\mathbb{Z}_2$ is $\phi \phi \phi \to \phi S$ where $S = h_1,h_2$. We list below the Feynman diagrams for illustration.

\begin{figure}[htpb!]{\centering 
   
 \includegraphics[scale =0.4, angle=0]{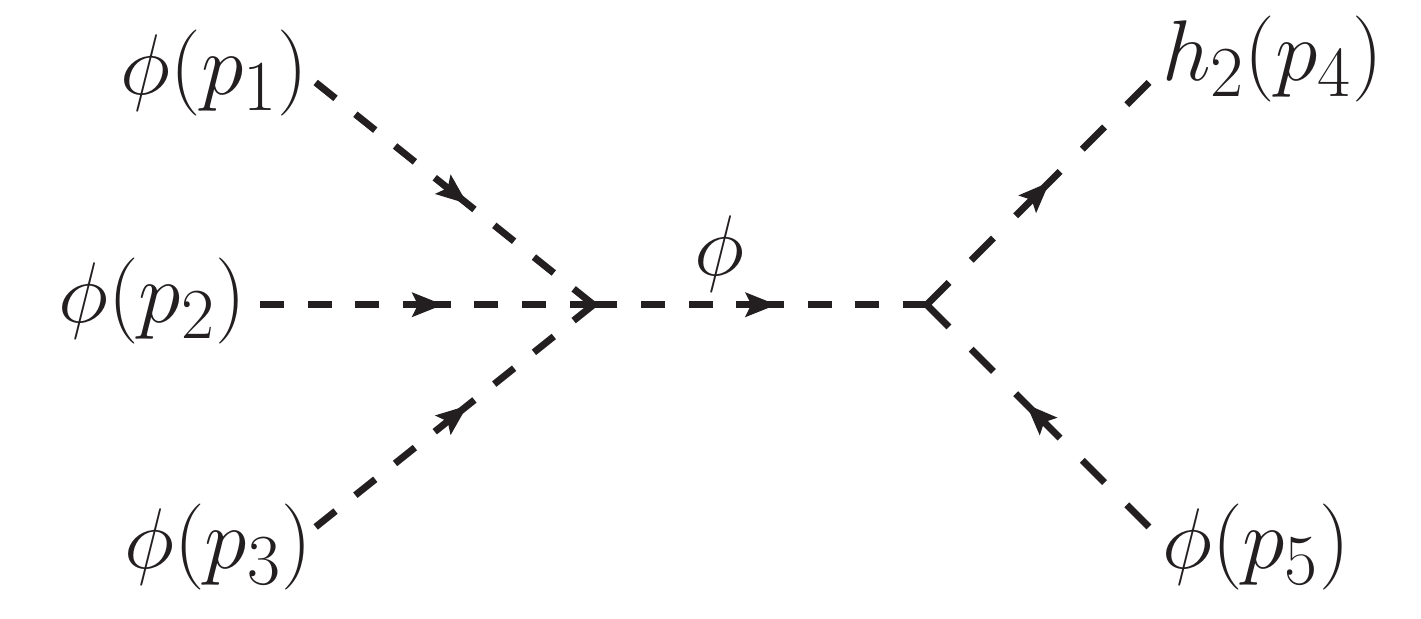}
   }
 \caption{$\phi$-mediated Feynman diagram responsible for annihilation of $\phi$.}
 \label{fig1}
 \end{figure}
 
  \begin{figure}[htpb!]{\centering 
   \subfigure[]{
 \includegraphics[scale =0.4, angle=0]{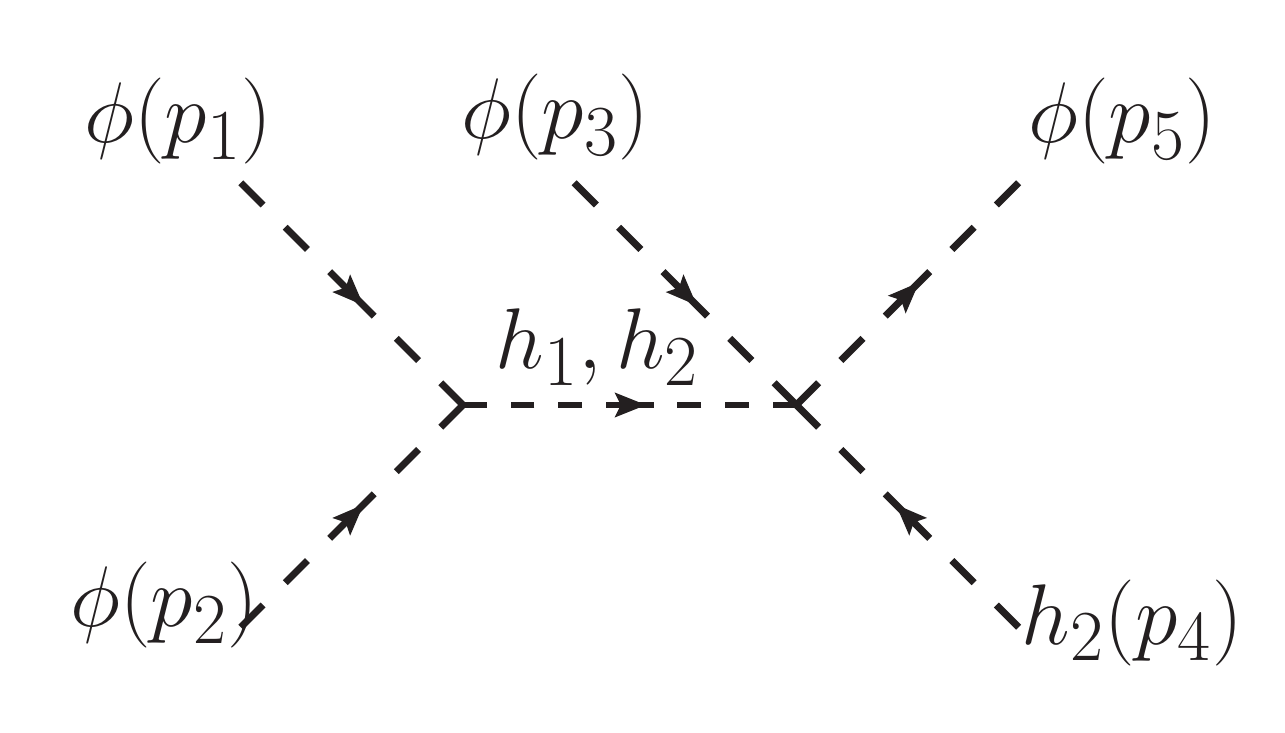}}
    \subfigure[]{
 \includegraphics[scale =0.4, angle=0]{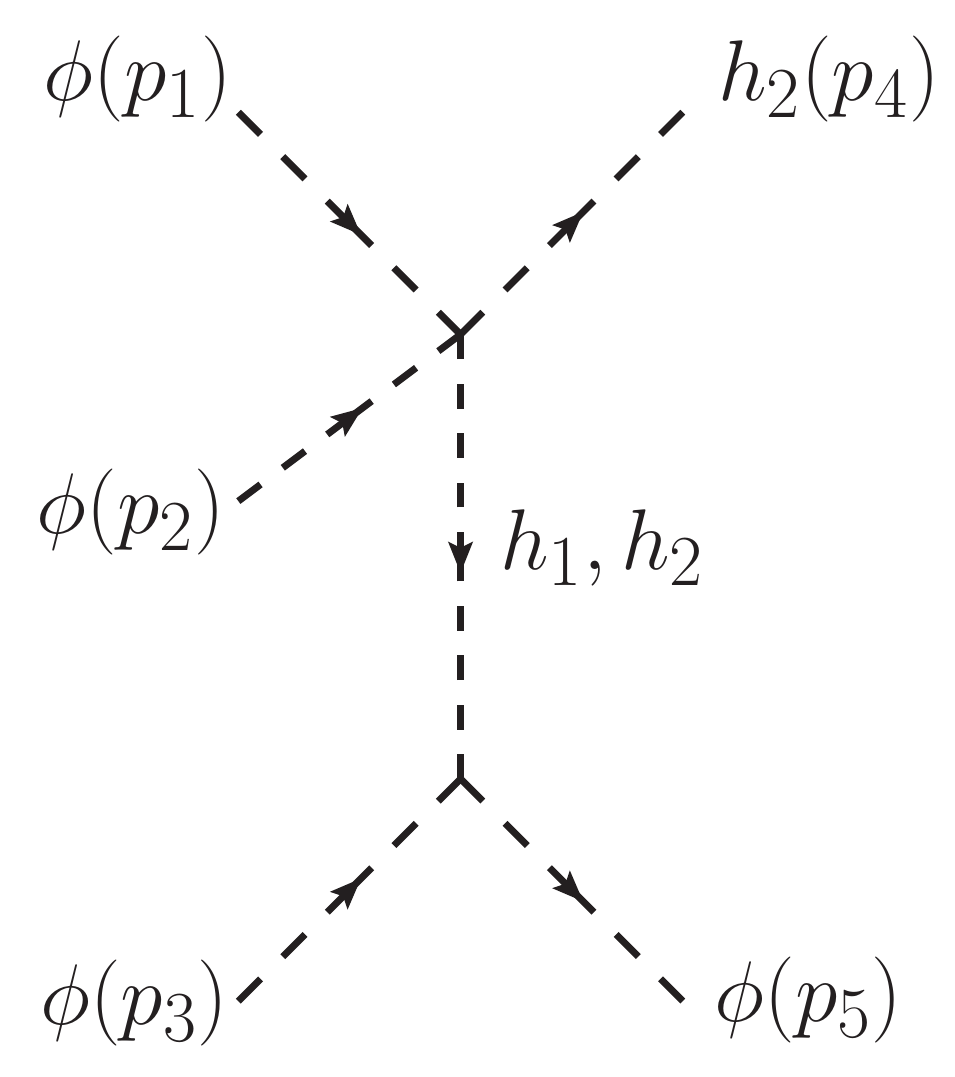}}
  }
 \caption{ Feynman diagrams show annihilation of $\phi$ mediated by $h_{1}$, $h_{2}$. }
 \label{fig2}
 \end{figure}
 
 \begin{figure}[htpb!]{\centering 
   \subfigure[]{
 \includegraphics[scale =0.4, angle=0]{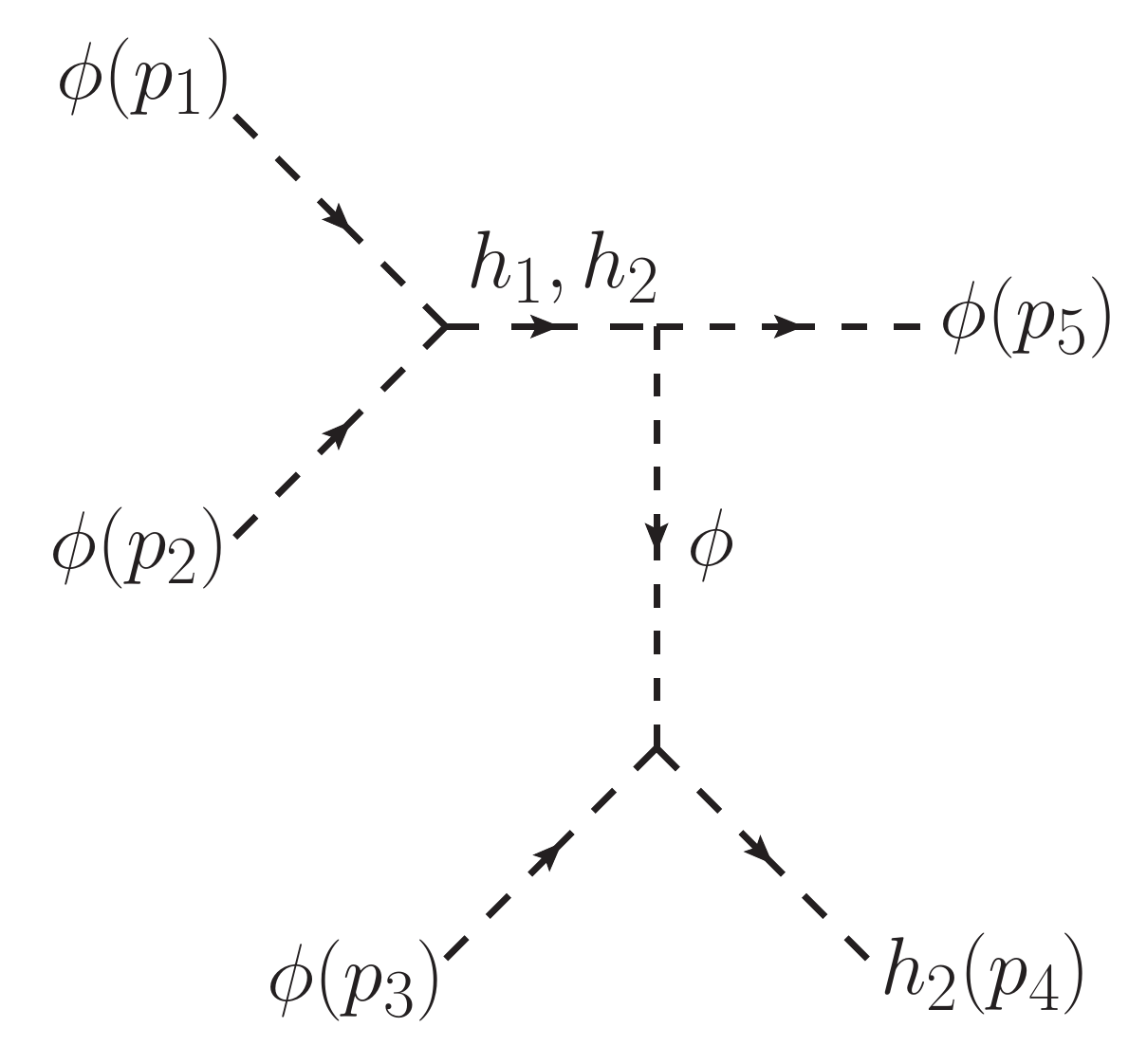}}  
 \subfigure[]{
 \includegraphics[scale =0.4, angle=0]{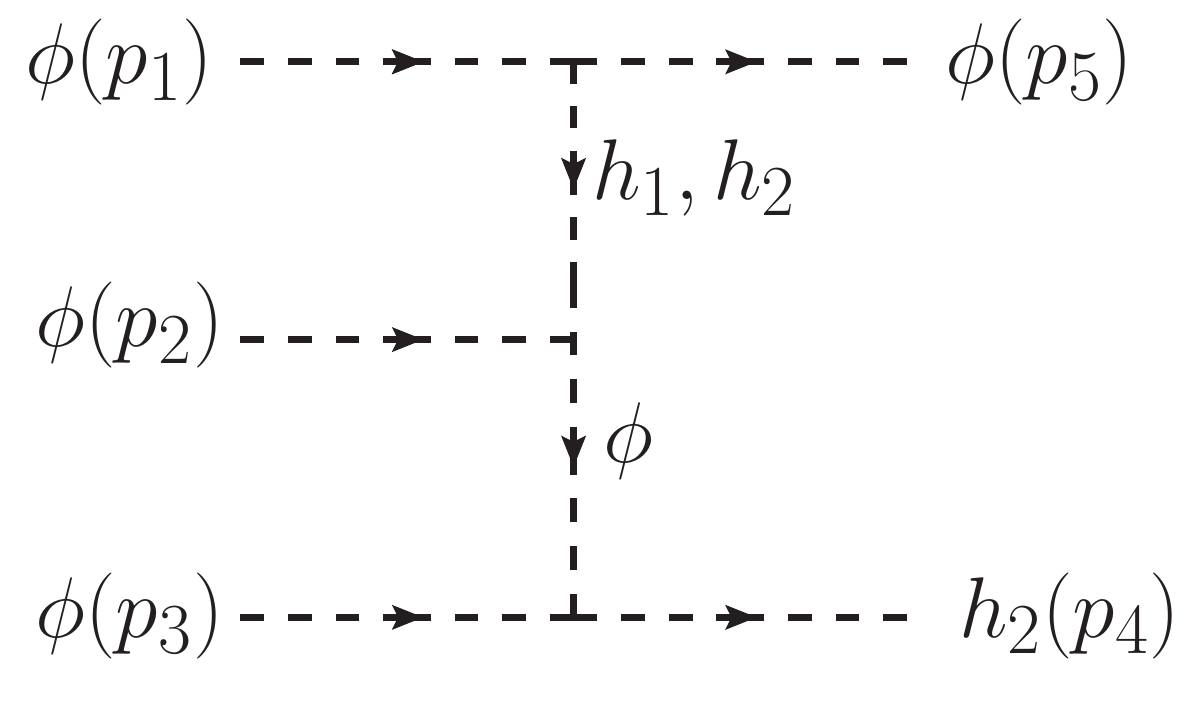}} \\
  \subfigure[]{
 \includegraphics[scale =0.4, angle=0]{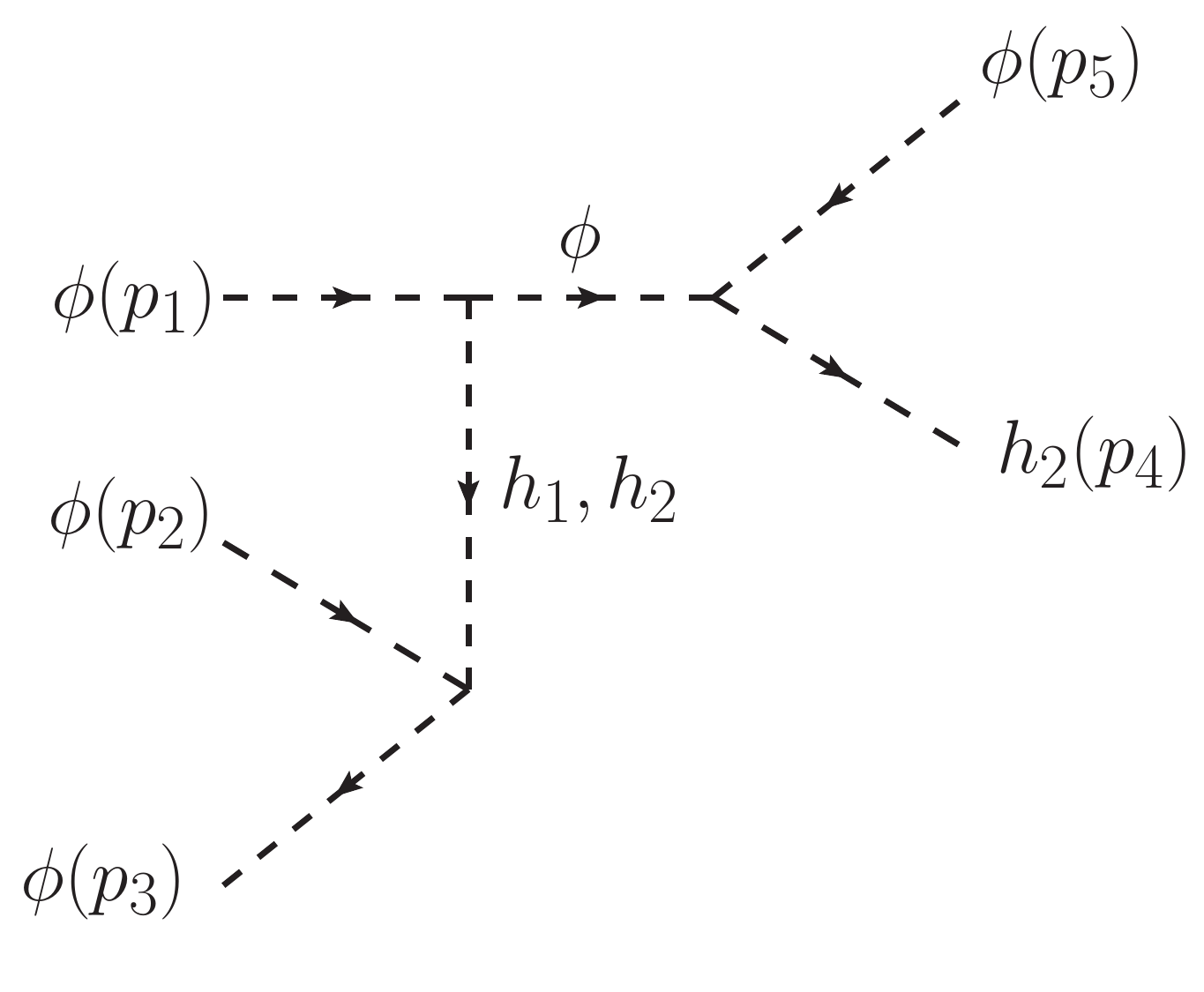}} 
 }
  \caption{ Annihilation diagrams  mediated by $h_{1}$, $h_{2}$ and $\phi$.}
 \label{fig3}
 \end{figure}
Figs.\ref{fig1}, \ref{fig2} and \ref{fig3}
show the $3 \to 2$ annihilation diagrams. 
We take the limit where sin$\theta$ and $\l_{13}$ are tiny. The tiny sin$\theta$ conforms with the constraints on $\delta-H$  mixing from direct search and Higgs signal strengths. The $\phi-\phi-h_1$ trilinear coupling becomes negligibly small in this limit while the $\phi-\phi-h_2$ coupling becomes $\simeq (\mu_{21} + \l_{12}v_\delta)$. Moreover, the limit entails that the $\phi$-mediated amplitude in Fig.\ref{fig1} is the dominant one. The following is its expression:
\bea
\mathcal{M} \simeq \frac{\l_{11}(\mu_{21} + \l_{12} v_\delta)}{(p_1 + p_2 + p_3)^2 - m^2_{\phi}} 
\simeq \frac{\l_{11}(\mu_{21} + \l_{12} v_\delta)}{8m_\phi^2}.
\eea
In the last step above, we have neglected the DM velocity, a justifiable assumption for non-relativistic DM. One can approximate $p_i \simeq (m_\phi,\vec{0})$ in that case. The evolution of the DM comoving number density $Y_\phi$ is described by the Boltzmann equation~\cite{Kolb:1990vq}
\bea
\frac{dY_{\phi}}{dx} = - \frac{x~s^{2}}{3\mathcal{H}} \langle \sigma v^2 \rangle_{\phi \phi \phi \rightarrow \phi h_2} \Big( Y^{3}_{\phi} - Y_{\phi} Y^{\text{eq}}_{\phi} Y^{\text{eq}}_{h_2} \Big), \label{beq}
\eea
where $s = \frac{2 \pi^{2}}{45}~ g_{*s} T^{3} $ is the entropy density and the Hubble parameter is $\mathcal{H} = \sqrt{\frac{8 \pi^{3} G g_{*} T^{4}}{90}}$\footnote{$g_{*}$ is the effective relativistic degrees of freedom.}. Besides, $Y^{\text{eq}}_P$ denotes the equilibrium comoving number density of a particle $P$ and 
$\langle \sigma v^2 \rangle_{\phi \phi \phi \rightarrow \phi h_2}$
is the thermally averaged $3 \to 2$ cross section~\cite{Pierre:2018man}. It can be related to $\sigma v^2 _{\phi \phi \phi \rightarrow \phi h_2}$ using modified Bessel functions $K_{n}(x)$ as

\bea
\braket{\sigma v^2}_{\phi \phi \phi \to \phi h_2}(T) &=&
\hspace{-2mm}\Bigg( \frac{K_1\big(\frac{m_\phi}{T}\big)}{K_2\big(\frac{m_\phi}{T}\big)} \Bigg)^3
(\sigma v^2)_{\phi \phi \phi \to \phi h_2},\nonumber \text{with} \\ \nonumber
\eea
\small

\bea
(\sigma v^2)_{\phi \phi \phi \to \phi h_2} &=& \frac{|\mathcal{M}|^2}{72 \pi m^3_\phi} \nonumber \\
 \times && \hspace{-6mm}\sqrt{\bigg(1 - \frac{5}{16}
\Big(\frac{m_{h_2}}{m_\phi}\Big)^2
+ \frac{1}{64}
\Big(\frac{m_{h_2}}{m_\phi}\Big)^4 \bigg)}. \label{sig_vsq}
\eea

Further, Eq.(\ref{beq}) becomes the following upon taking $g_{*s} \simeq g_{*}$.
\bea
\frac{dY_{\phi}}{dx} &=& - 0.116~ g^{3/2}_{*}~ M_{\text{Pl}} \frac{m^{4}_{\phi}}{x^{5}} ~  \langle \sigma v^2 \rangle_{\phi \phi \phi \rightarrow \phi h_2}\nonumber \\
&&~~~~~~~~~~~~~~~~~~   \times \Big( Y^{3}_{\phi} - Y_{\phi} Y^{\text{eq}}_{\phi} Y^{\text{eq}}_{h_2} \Big). \label{beq2}
\eea
We present an approximate analytical solution to Eq.(\ref{beq2}) to gain insight on the dynamics. We follow here the approach adopted in~\cite{Bhattacharya:2019mmy}. We take $B \equiv 0.116~ g^{3/2}_{*}~ M_{\text{Pl}}~m^{4}_{\phi} ~\langle \sigma v^2 \rangle_{\phi \phi \phi \rightarrow \phi h_2}$ and rewrite Eq.(\ref{beq2})
using $\Delta = Y_{\phi} - Y^{\text{eq}}_{\phi}$ as
\bea
\frac{d(\Delta + Y^{\text{eq}}_{\phi})}{dx} &=& - \frac{B}{x^{5}} \Big[(\Delta + Y^{\text{eq}}_{\phi})^{3} \nonumber \\
&- & (\Delta + Y^{\text{eq}}_{\phi}) ~Y^{\text{eq}}_{\phi}~ Y^{\text{eq}}_{h_2} \Big].\label{del_yeq}
\eea
Before freeze-out, i.e. for $1 \ll x \leq x_{f}$ ($x_f$ denotes freeze out of DM), $\Delta \ll Y^{eq}_{\phi}$ and $\frac{d\Delta}{dx} \rightarrow 0$ and near freeze-out, i.e. for $x \simeq x_{f}$, so $\Delta(x_{f}) = c Y^{\text{eq}}_{\phi} (x_{f})$. One then writes
\bea
\frac{dY^{\text{eq}}_{\phi}(x_{f})}{dx} &=&  - \frac{B}{x^{5}} \Big[ (c+1)^{3} (Y_{\phi}^{\text{eq}})^{3} (x_{f}) \nonumber \\
&- &   (c+1)~(Y_{\phi}^{\text{eq}})^{2} (x_{f}) (Y_{h_2}^{\text{eq}}) (x_{f}) \Big].\label{beq_approx}
\eea
Using the equilibrium distributions
\bea
Y^{\text{eq}}_{\phi} (x) &=& 0.145~\Big(\frac{g_{\phi}}{g_{*}} \Big)~ x^{3/2} ~e^{-x}, \nonumber \\
Y^{\text{eq}}_{h_2} (x) &=& 0.145 ~\Big(\frac{g_{h_2}}{g_{*}} \Big)~\Big(\frac{m_{h_2}}{m_{\phi}} \Big) ~x^{3/2}~ e^{- \frac{m_{h_2}}{m_{\phi}} x},
\eea
Eq.(\ref{beq_approx}) takes the form
\bea
x^2_f - \frac{3}{2} x_f &=& (0.145)^2 B (c + 1) \Big(\frac{g_\phi}{g_*}\Big)^2 e^{-2 x_f}\nonumber \\
&\times & \Big[(c+1)^2 - \Big(\frac{m_{h_2}}{m_\phi} \Big)^{3/2} e^{-\Big(\frac{m_{h_2}}{m_\phi} + 1 \Big)x_f}  \Big].\label{xff}
\eea

Eq.\ref{xff} can be iteratively solved for $x_f$. Now $Y^{\text{eq}}_\phi \ll  \Delta$ for $x \gg x_f$. Eq.\ref{del_yeq} can then be integrated from the freeze-out epoch $x=x_f$ to a later epoch $x \to \infty$ as
\bea
\int_{\Delta(x_f)}^{\Delta(x \to \infty)} \frac{d \Delta}{\Delta^3} = -B \int_{x_f}^{\infty} dx~\frac{1}{x^5}
\eea
to finally yield $Y_\phi(x \to \infty) = x_f^2 \bigg[ \frac{2}{0.12 M_{\text{Pl}} m^4_\phi \braket{\sigma v^2}} \bigg]^{1/2}$. The final relic abundance is related to the comoving density $Y_\phi(x \to \infty)$ as
$\Omega h^2 = 2.752 \times 10^8~m_\phi Y_\phi(x \to \infty)$ \cite{Kolb:1990vq,Bhattacharya:2016ysw}.

\section{First order phase transitions and numerical analysis}\label{sfopt}

We are interested in cosmological phase transitions in the direction of the field $\delta$ given its crucial role in the generation of the DM relic through the $3 \to 2$ transitions. The tree-level potential in the $\delta$-direction looks like $V_0(\delta_{\text{cl}}) = \frac{\mu_{\delta}^2}{2} \delta_{\text{cl}}^2 + \frac{\mu_{22}}{3!} \delta_{\text{cl}}^3 +\frac{\lambda_{22}}{4!} \delta_{\text{cl}}^4$. Here $\delta_{\text{cl}}$ denotes the classical rolling field and is distinct from the degree of freedom $\delta^\prime$. Next, the one-loop Coleman-Weinberg correction to the tree-level potential reads
\bea
\Delta V_{\rm CW}(\delta_{\text{cl}}) &=& \hspace{-2mm}\frac{1}{64 \pi^2} \hspace{-6mm} \sum_{a=\phi,\delta^\prime,h,G^+,G^0} \hspace{-4mm} n_a  \bigg[ m^4_a(\delta_{\text{cl}}) \Big( \text{log} \frac{m_i^2(\delta_{\text{cl}})}{m_a^2(v_\delta)} - \frac{3}{2}\Big)\nonumber \\
& &\hspace{2.8cm}+ 2 m^2_a(v_\delta)m^2_a(\delta_{\text{cl}})\bigg]. \label{VCW}
\eea
The sum in Eq.(\ref{VCW}) runs over the scalars of the theory since these are the fields that develop $\delta_{\text{cl}}$-dependent masses. Also, $n_a$ denotes number of degrees of freedom of the $a$th scalar. One finds $n_\phi=n_\delta=n_h=n_{G^0} = 1$ while $n_{G^+} = 2$. The field-dependent masses are
\besub
\bea
m^{2}_{\phi} (\delta_{\text{cl}}) &=& \mu_{\phi}^2 + \frac{\lambda_{12} \hspace{0.5mm} \delta_{\text{cl}}^2}{2} + \mu_{21}\hspace{0.5mm} \delta_{\text{cl}}, \\
m^{2}_{\delta} (\delta_{\text{cl}}) &=& \mu_{\delta}^{2} + \frac{\lambda_{22} \hspace{0.5mm} \delta_{\text{cl}}^2}{2}  + \mu_{22} \hspace{0.5mm} \delta_{\text{cl}},  \\ 
m^{2}_{h} (\delta_{\text{cl}}) &=& m^{2}_{G^0}(\delta_{\text{cl}}) = m^{2}_{G^+} (\delta_{\text{cl}}) ,\nonumber \\
&=& \mu_h^2  + \frac{\lambda_{23}\hspace{0.5mm} \delta_{\text{cl}}^2}{2} + \mu_{23} \hspace{0.5mm}\delta_{\text{cl}}.
\eea 
\eesub
The one-loop correction to the scalar potential induced in presence of $T \neq 0$ reads
\bea
\Delta V_1(\delta_{\text{cl}},T) &=& \frac{T^4}{2 \pi^2} \hspace{-4mm}\sum_{a=\phi,\delta^\prime,h,G^+,G^0} \hspace{-6mm} n_a J_B[m^2_a(\delta_{\text{cl}})/T^2].
\eea
Here, the function $J_{B}(\frac{m^2}{T^2})$ is defined as $J_{B}(\frac{m^2}{T^2}) = \int_0^\infty dx~x^2~\text{log}\big[1 - e^{-\sqrt{x^2 + \frac{m^2}{T^2}}} \big]$. It can be approximated in the high temperature limit as
\bea
J_{B}\Big(\frac{m^2}{T^2}\Big) &\simeq & - \frac{\pi^{4}}{45} + \frac{\pi^{2}}{12}~\frac{m^2}{T^2} - \frac{\pi}{6}~\frac{m^3}{T^3}\nonumber \\
&& - \frac{1}{32}~\frac{m^4}{T^4}~\text{ln}\Big(\frac{m^2}{a_{B}T^2}\Big),\label{JB_highT}
\eea

where $a_{B} \equiv 16\pi^2~e^{\frac{3}{2}-2\gamma_{E}} $. Further, the infrared effects are included using the daisy remmuation technique~\cite{Weinberg,Dolan,Kirzhnits:1976ts,Carrington:1991hz,Gross,Fendley:1987ef,Kapusta,Arnold:1992rz,Parwani:1991gq}. More precisely, the Arnold-Espinosa prescription~\cite{Arnold:1992rz} in adopted in this study. That is
\bea
\Delta V_{\text{ring}}(\delta_{\text{cl}},T) &=&  -\frac{T}{12 \pi}  \sum_{\substack{ a=\phi,\delta^\prime, \\ h,G^+,G^0}} \hspace{-4mm} \Big[(m_a^2(\delta_{\text{cl}},T))^{3/2}  -  (m_a^2(\delta_{\text{cl}}))^{3/2} \Big]. \nonumber \\
\eea
The total thermal scalar potential as a function of $\delta_{cl}$ and $T$ is the sum of the individual components. That is
\bea
V_{\text{total}}(\delta_{\text{cl}},T) &=& V_0(\delta_{\text{cl}}) + \Delta V_{\rm CW}(\delta_{\text{cl}}) \nonumber \\
& +& \Delta V_1(\delta_{\text{cl}},T) + \Delta V_{\text{ring}}(\delta_{\text{cl}},T).
\eea
We attempt to understand $V_{\text{total}}(\delta_{\text{cl}},T)$ more intuitively with the aid of appropriate approximations. The thermal potential can be cast as a polynomial in $\delta_{\text{cl}}$ using the high temperature approximation for $J_B(\frac{m^2}{T^2})$ given in Eq.(\ref{JB_highT}) and discarding terms that do not contain $\delta_{\text{cl}}$. One then straightforwardly derives
\bea
V_{\text{total}}(\delta_{\text{cl}},T) & \simeq & \Big[ \frac{\mu_{\delta}^{2}}{2} +  \frac{1}{48}(\lambda_{12}+\lambda_{22}+\lambda_{23})~T^{2}\Big]~\delta_{\text{cl}}^{2}  \nonumber \\  &-& \frac{1}{24\sqrt{2} \pi}\Big[ \Big(\lambda^{3/2}_{12} + \lambda^{3/2}_{22} + \lambda^{3/2}_{23}  \Big)~ T - \frac{\mu_{22}}{6} \Big]~\delta_{\text{cl}}^{3} \nonumber \\  &+& \frac{\bar{\lambda}(T)}{4} \delta_{\text{cl}}^{4}. 
\eea
where, $\bar{\lambda}(T) = \frac{\lambda_{22}}{6} + \frac{1}{64 \pi^2}\Big[\lambda^{2}_{12} ~\text{ln} \Big(\frac{a_{B}~T^{2}}{e^{3/2}~ m^{2}_{\phi}(v_{\delta})}\Big) + \lambda^{2}_{22} ~\text{ln} \Big(\frac{a_{B}~T^{2}}{e^{3/2}~ m^{2}_{\delta}(v_{\delta})}\Big) + \lambda^{2}_{23} ~\text{ln} \Big(\frac{a_{B}~T^{2}}{e^{3/2}~ m^{2}_{h}(v_{\delta})}\Big) \Big]$. \label{VT_poly}
The form in Eq.(\ref{VT_poly}) resembles the SM formula in~\cite{Quiros:1999jp} and therefore permits an analytical treatment of the thermal evolution. We identify certain temperature thresholds in Eq.(\ref{VT_poly}). An extrema at $\delta_{\text{cl}} = 0$ is readily identified. Moreover, this extrema is a minima (maxima) for $T > T_0$ ($T < T_0$) with $T_0 = \sqrt{-24 \mu^2_\delta/(\l_{12}+\l_{22}+\l_{23})}$. An inflection point is present for $T=T_1$ is encountered.
The temperature $T_1$ can be solved from $\Big[ \Big(\lambda^{3/2}_{12} + \lambda^{3/2}_{22} + \lambda^{3/2}_{23}  \Big)~ T_1 - \frac{\mu_{22}}{6} \Big]^2 = 1152 \pi^2 \bar{\l}(T_1) \Big[ \mu_{\delta}^{2} +  \frac{1}{24}(\lambda_{12}+\lambda_{22}+\lambda_{23})~T_1^{2}\Big]$. For $T_0 < T < T_1$, a maxima and a minima appear at $\delta_{\text{cl}} = \delta_{\text{max}}(T),\delta_{\text{min}}(T)$ in addition to the minima at $\delta_{\text{cl}}=0$. Thus, the $T_0 < T < T_1$ temperature band is the most interesting from the perspective of first order phase transitions on account of the coexisting minima.  The critical temperature $T_c$ for this case is the temperature at which the minima at $\delta_{\text{cl}}=0,\delta_{\text{min}}$ are degenerate, i.e., $V_{\text{total}}(0,T_c)=V_{\text{total}}(\delta_{\text{min}}(T_c),T_c)$. A strong first order phase transition is identified by $\frac{\delta_{\text{min}}(T_c)}{T_c} \geq 1$.

\begin{figure*}[t!]{\centering 
 \includegraphics[scale =0.40, angle=0]{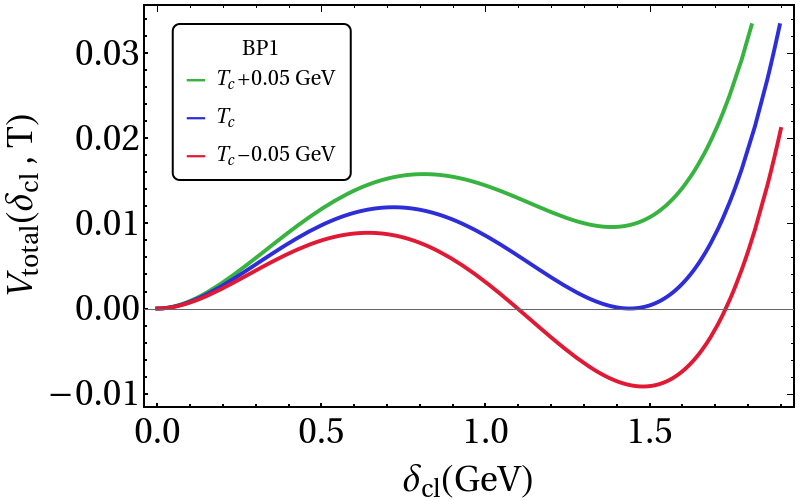}  
 \includegraphics[scale =0.40, angle=0]{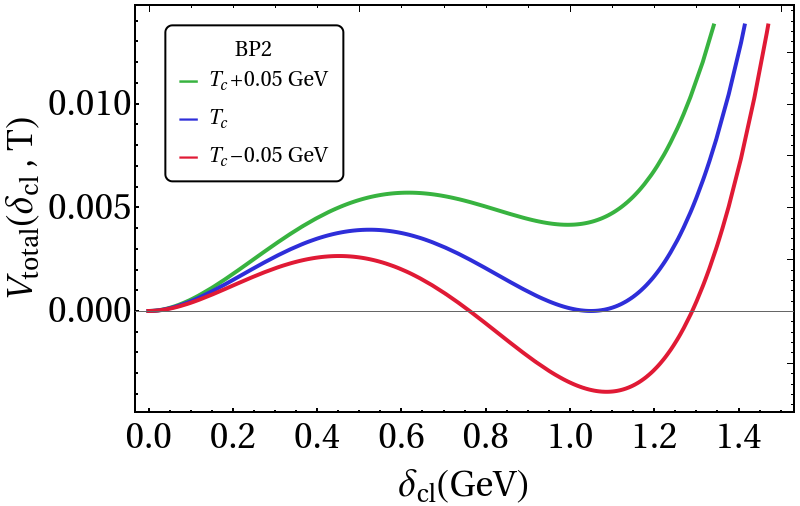} 
 }
  \caption{The extrema of the thermal potential around $T=T_c$ for BP1 and BP2. The color coding is shown in the legends.}
 \label{Vtotal}
 \end{figure*}

We put forth two benchmark points in Table \ref{table:BP}. The independent model parameters and the corresponding critical temperatures are listed therein. These benchmarks also predict relic densities within the Planck band. The shape of $V_{\text{total}}(\delta_{\text{cl}},T)$ around the crtical temperatures is shown in Fig.\ref{Vtotal}.
Both BP1 and BP2 correspond to sub-GeV critical temperatures. One also inspects in Fig.\ref{Vtotal} that $\delta(T_c) \gtrsim$ 1 GeV for both thereby signalling a SFOPT.

%
%

\begin{table*}[t!]
	\centering
	\resizebox{17cm}{!}{
	\begin{tabular}{|c|c|c|c|c|c|c|c|c|c|c|c|c|c|}
		\hline 
  & $\lambda_{11}$ &  $\lambda_{12}$ & $\lambda_{22}$ & $\lambda_{23}$ &  $\mu_{22}$ (GeV) & $ m_{h_2}$ (GeV)  & $m_{\phi}$ (GeV) & $v_{\delta}$ (GeV) & $\mu_{21}$ (GeV) & $T_{c}$ (GeV)& $\frac{\delta(T_c)}{T_c}$ & $\Omega_{\text DM} h^{2}$ \\ \hline
  
BP1 & 0.1 & 4.69202 & 0.0060999 & 0.00001 & -0.208174 & 0.0647581 & 0.050095 & 66.8 & 0.756 & 0.975268 & 1.47608 & 0.114796 \\ \hline

BP2  & 0.5 & 5.39401 & 0.0060999 & 0.00001 & -0.144512 & 0.116026 & 0.0830917 & 45.3 & 0.586 & 0.717489 &1.46328 &0.114441 \\ \hline
   \end{tabular}}
	\caption{Benchmark model parameters along with the predicted relic densities and the crtical temperatures. }

	\label{table:BP}
\end{table*}

\begin{figure*}[t!]{\centering 
   \subfigure[]{
 \includegraphics[scale =0.35, angle=0]{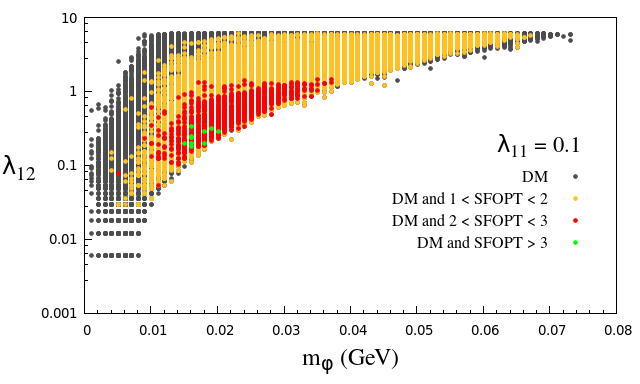}}  
 \subfigure[]{
 \includegraphics[scale =0.35, angle=0]{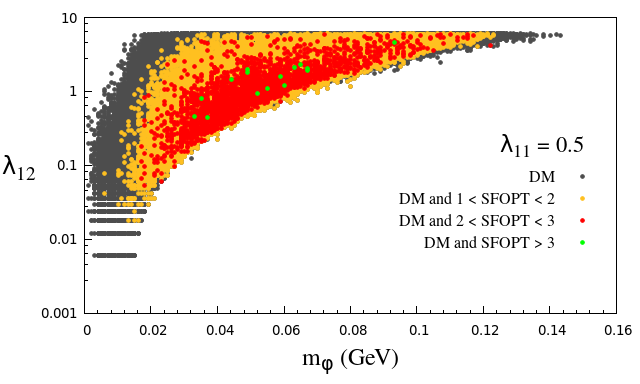}} \\
  \subfigure[]{
 \includegraphics[scale =0.35, angle=0]{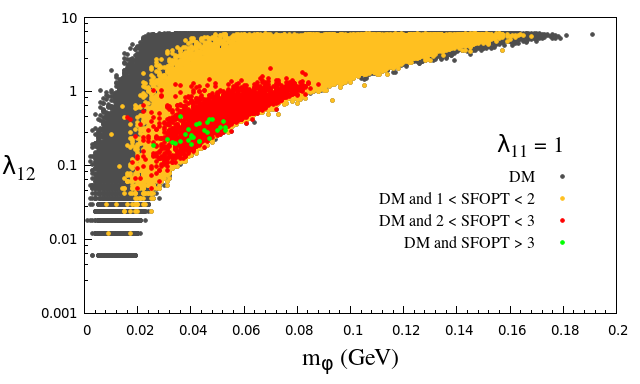}} 
 }
  \caption{Parameter points in the $m_\phi-\l_{12}$ plane allowed by $\Omega h^2$ and SFOPT criteria for $\l_{11}$ = 0.1 (top left), 0.5 (top right) and 1 (bottom). The color coding is described in the legends.}
 \label{mphi_l12}
 \end{figure*}

\begin{table*}[t!]
	\centering
	\resizebox{14cm}{!}{
	\begin{tabular}{|c|c|c|c|c|c|c|}
		\hline 
    & $T_{n}$ (GeV) & $\alpha$ & $\beta/H$ & $f_{\text{col}}$ (Hz) & $f_{\text{sw}}$ (Hz) & $f_{\text{tur}}$ (Hz)  \\ \hline 
    
 BP1 &  0.821609  &  0.00931039 &  1119.22 & 4.16229 $\times 10^{-5}$ & 2.71898 $\times 10^{-4}$ & 3.86382 $\times 10^{-4}$\\ \hline 
 
  BP2 &  0.606572 &  0.009944 & 1149.43 &  3.15148 $\times 10^{-5}$ & 2.0544 $\times 10^{-4}$ & 2.91941 $\times 10^{-4}$\\ \hline 

   \end{tabular}}
	\caption{The nucleation temperature and other GW parameters corresponding to BP1 and BP2.}
	\label{alpha_beta}
\end{table*}
A combined analysis of the relic density and SFOPT is in order. We first fix $\l_{13} = \l_{23} = 10^{-5},~\l_{11}=0.1,~0.5,~1$ and vary the rest of the parameters as $|\sin\theta| < 0.001,~|\l_{12}| < 6,~|\l_{22}| < 6,~|\mu_{21}| < 1~\text{GeV},~ 1~\text{MeV} < m_\phi < 1~\text{GeV},~ 1~\text{MeV} < m_{h_2} < 2 m_\phi$. It can be noted that the small values taken for sin$\theta$ and $\l_{13}$ render the $\phi-\phi-h_{1(2)}$ interactions appropriately small so as to conform with possible stringent direct detection bounds in the near future. The small sin$\theta$ also implies safety from the Higgs signal strength constraints and the direct search constraints involving $h_2$. The full thermal potential is analysed using the publicly available tool \texttt{PhaseTracer} \cite{Athron:2020sbe}. We select the parameter points that lead to (a) $0.1118 < \Omega h^2 < 0.1280$~\cite{Planck:2018vyg} and (b) $\frac{\delta_{\text{min}}(T_c)}{T_c} > 1$. The results are shown as plots in the $m_{\phi}-\l_{12}$ plane in Fig.\ref{mphi_l12}.
 
Fig.\ref{mphi_l12} shows a $\mathcal{O}$(10)-$\mathcal{O}$(100)(MeV) mass range for the DM thereby corroborating the results of~\cite{Mohanty:2019drv}. The correlation among $m_\phi$, $\l_{11}$ and $\l_{12}$ can be understood from $(\sigma v^2)_{\phi \phi \phi \to \phi h_2} \sim \frac{\l^2_{11}(\mu_{21}+\l_{12}v_\delta)^2}{m^7_\phi}$ in Eq.(\ref{sig_vsq}). The maximum   
allowed value of $m_\phi$ increases with increasing $\l_{11}$. And this is concurred by Fig.\ref{mphi_l12} where one inspects that for the aforesaid variation of $\l_{12}$, changing $\l_{11}=0.1$ to $\l_{11}=1$ takes one from $m_\phi \lesssim$ 70 MeV to $m_\phi \lesssim$ 190 MeV. Demanding a SFOPT restricts the parameter space further.
 \begin{figure*}[htpb!]{\centering 
 \includegraphics[scale =0.40, angle=0]{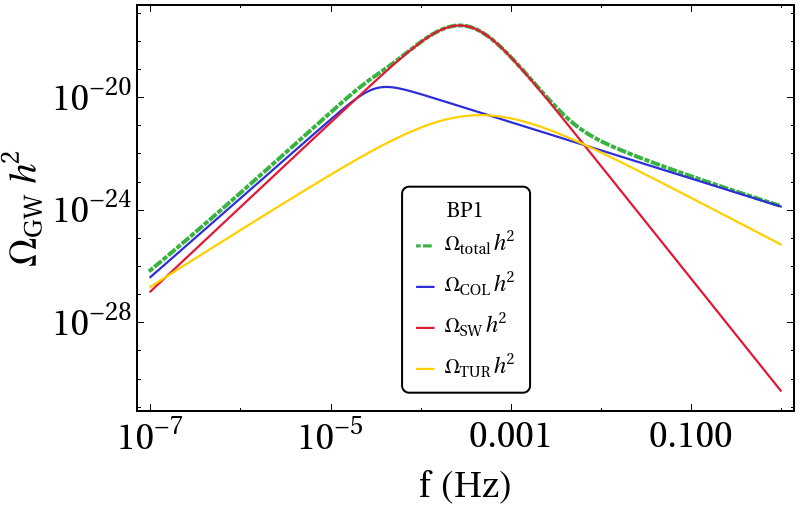}\vspace{8mm}
 \includegraphics[scale =0.40, angle=0]{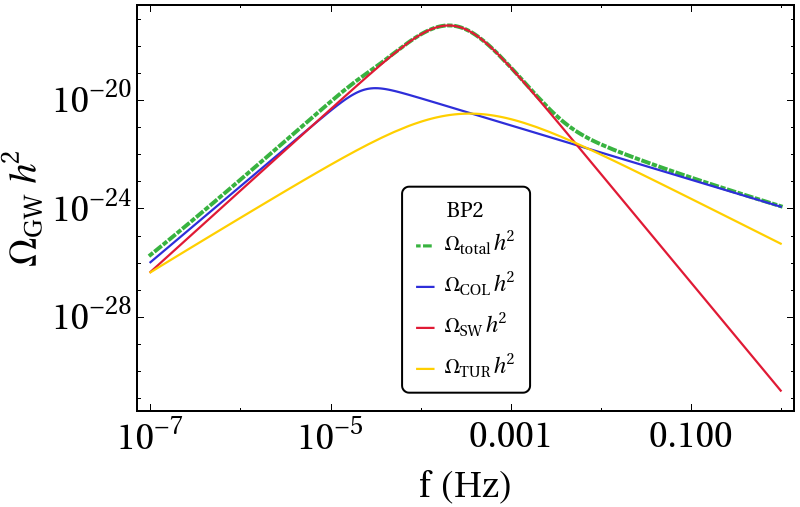} 
 }
  \caption{The nucleation temperature and other GW parameters corresponding to BP1 and BP2.}
 \label{f:OmegaGW}
 \end{figure*}

We now discuss the GW spectrum of such a scenario and a few pertinent quantities are therefore introduced. The parameter $\beta$ is defined as
\bea
\frac{\beta}{H} &=&\bigg[ T \frac{d}{d T}\Big(\frac{S_3}{T} \Big) \bigg]_{T_n}. \label{beta_expr}
\eea  
Here, $T_n$ is the nucleation temperature and $S_3$ is the Euclidean action in three dimensions. Next, we define $\Delta V_{\text{tot}}(T) \equiv V_{\text{tot}}(0,T) - V_{\text{tot}}(\delta_{\text{min}}(T),T)$ which measures the difference in the depths of the potential at the two vacua at a temperature $T$. The energy budget of the phase transition during the bubble nucleation is then given by
\bea
\epsilon &=& \Delta V_{\text{tot}}(T_n) - \bigg[T \frac{d \Delta V_{\text{tot}}(T)}{d T}\bigg]_{T_n}.
\eea

One subsequently defines $\alpha = \frac{\epsilon}{\rho_n}$, where $\rho_n = \frac{g_*\pi^2}{30}T_n^4$,
is the energy density during nucleation. The sources of GWs associated with SFOPT are  bubble collision~\cite{Kosowsky,Turner,Huber_2008,
Watkins,Marc,Caprini_2008}, sound waves~\cite{Hindmarsh,Leitao:2012tx,Giblin:2013kea,Giblin:2014qia,Hindmarsh_2015}, and, magnetohydrodynamic turbulence~\cite{Chiara,Kahniashvili,Kahniashvili:2008pe,Kahniashvili:2009mf,Caprini:2009yp}.
The GW amplitudes from these sources are functions of the frequency. These amplitudes attain their maximum values at certain ``peak frequencies'' that we denote as $f_{\text{col}},f_{\text{sw}}$ and $f_{\text{tur}}$. The expressions for these frequencies and the corresponding GW amplitudes is given in the Appendix \ref{app}.

The $\a$ and $\b$ parameters are crucial in deciding the strength of the GW signals stemming from a SFOPT.
Table~\ref{alpha_beta} displays $T_n,\alpha,\beta$ for the chosen benchmarks. Bubble nucleation at the sub-GeV temperatures shown can be understood as an artefact of the sub-GeV DM masses involved. This is an important observation of our analysis. We also inspect in table~\ref{alpha_beta} that $\a$ and $\beta/H$ at nucleation are respectively $\mathcal{O}(10^{-3})$ and $\mathcal{O}(10^3)$ for the BPs. Such ball-park values entail that the magnitude of the total GW amplitude peaks at $\mathcal{O}(10^{-18})$, as is seen in Fig.\ref{f:OmegaGW}. And the corresponding
frequencies are $\sim \mathcal{O}(10^{-4})$ Hz. This is expected since the total GW amplitude is dominated by the sound wave component that has $f_{\text{sw}} \sim \mathcal{O}(10^{-4})$ Hz. Overall, an  
$\Omega_{\text{GW}}h^2 \sim \mathcal{O}(10^{-18})$ with peak at $f \sim \mathcal{O}(10^{-4})$ Hz is beyond the reach of the proposed GW detectors. However, u-DECIGO~\cite{Kudoh:2005as} is expected to detect GW signals of a similar magnitude, albeit at a different frequency range. In all, a future detector with higher sensitivity in the $\sim \mathcal{O}$(0.1) mHz regime can potentially detect GW signals of a sub-GeV DM species exhibiting number changing annihilations.

\section{Conclusions}\label{conclu}

In this study, we look at possible GW signals originating from a SIMP. The SIMP species freezes-out via $3 \to 2$ number changing processes that are in turn driven by another scalar $\delta$. The observed relic density is attained for sub-GeV DM masses. We have studied the shape of the thermal potential along $\delta$ and identified the region of the parameter space permitting a SFOPT. For a few representative benchmarks, we have then computed the GW amplitude as a function of frequency. Our analysis reveals that a futuristic detector with a sensitivity $\Omega_{\text{GW}}h^2 \sim 10^{-18}$ in the mHz regime can detect GWs stemming from such a SIMP framework.

\section{Appendix}\label{app}

The peak frequencies corresponding to bubble collisions, sound waves and turbulence are expressed as 
\besub
\bea
f_\text{col} &=& 1.65 \times 10^{-5} 
\Big( \frac{0.62}{1.8 - 0.1 v_b + v^2_b} \Big) \nonumber \\
&&
 \frac{\beta}{H} \Big(\frac{T_n}{100}\Big) \Big(\frac{g_*}{100}\Big)^{\frac{1}{6}}, \\
f_\text{sw} &=& 1.9 \times 10^{-5} 
\Big( \frac{1}{v_b} \Big) \frac{\beta}{H} \Big(\frac{T_n}{100}\Big) \Big(\frac{g_*}{100}\Big)^{\frac{1}{6}}, \\
f_\text{tur} &=& 2.7 \times 10^{-5} 
\Big( \frac{1}{v_b} \Big) \frac{\beta}{H} \Big(\frac{T_n}{100}\Big) \Big(\frac{g_*}{100}\Big)^{\frac{1}{6}}.
\eea
\eesub

%
%

We express below the GW amplitudes from the three sources as a function of frequency $f$.
\bea
\Omega_{\text{coll}}(f) &=&  1.67 \times 10^{-5} \Big( \frac{\beta}{H} \Big)^{-2}\Big(\frac{0.11 v^3_b}{0.42 + v^2_b}\Big)
\Big( \frac{\kappa \a}{1 + \a} \Big)^2 \nonumber \\
&&
 \Big(\frac{g_*}{100}\Big)^{-\frac{1}{3}} 
\Big( \frac{3.8 (f/f_{\text{coll}})^{2.8}}{1+2.8(f/f_{\text{coll}})^{3.8}} \Big)
\eea
\bea
\Omega_{\text{sw}}(f) &=&  2.65 \times 10^{-6} \Big( \frac{\beta}{H} \Big)^{-2} v_b
\Big( \frac{\kappa_v \a}{1 + \a} \Big)^2  \nonumber \\
&&
 \Big(\frac{g_*}{100}\Big)^{-\frac{1}{3}} 
\Big(\frac{f}{f_{\text{sw}}} \Big)^3
\Big( \frac{7}{4+3(f/f_{\text{sw}})^{2}} \Big)^2,
\eea
\bea
\Omega_{\text{tur}}(f) &=&  3.35 \times 10^{-4} \Big( \frac{\beta}{H} \Big)^{-2} v_b
\Big( \frac{\epsilon \kappa_v \a}{1 + \a} 
 \Big)^{1.5} \nonumber \\
&&
  \Big(\frac{g_*}{100}\Big)^{-\frac{1}{3}} 
\Big(\frac{f}{f_{\text{tur}}} \Big)^3
\frac{(1 + f/f_{\text{tur}})^{-11/3}}{1+ 8\pi f/h_s}. 
\eea

\section{Acknowledgements}

NC acknowledges support from DST,
India, under grant number IFA18-PH214 (INSPIRE Faculty Award). HR is supported by
the Science and Engineering Research Board, Government of India, under the agreement
SERB/PHY/2016348 (Early Career Research Award). TS acknowledges the support from
the Dr. D. S. Kothari Postdoctoral scheme No. PH/20-21/0163.

\bibliographystyle{JHEP}
\bibliography{refer1}

\providecommand{\href}[2]{#2}\begingroup\raggedright\begin{thebibliography}{10}

\bibitem{Rubin:1970zza}
V.~C. Rubin and W.~K. Ford, Jr., \emph{{Rotation of the Andromeda Nebula from a
  Spectroscopic Survey of Emission Regions}},
  \href{http://dx.doi.org/10.1086/150317}{\emph{Astrophys. J.} {\bf 159} (1970)
  379--403}.

\bibitem{Bertone:2004pz}
G.~Bertone, D.~Hooper and J.~Silk, \emph{{Particle dark matter: Evidence,
  candidates and constraints}},
  \href{http://dx.doi.org/10.1016/j.physrep.2004.08.031}{\emph{Phys. Rept.}
  {\bf 405} (2005) 279--390}, [\href{http://arxiv.org/abs/hep-ph/0404175}{{\tt
  hep-ph/0404175}}].

\bibitem{Hu:2001bc}
W.~Hu and S.~Dodelson, \emph{{Cosmic Microwave Background Anisotropies}},
  \href{http://dx.doi.org/10.1146/annurev.astro.40.060401.093926}{\emph{Ann.
  Rev. Astron. Astrophys.} {\bf 40} (2002) 171--216},
  [\href{http://arxiv.org/abs/astro-ph/0110414}{{\tt astro-ph/0110414}}].

\bibitem{WMAP:2012nax}
{\scshape WMAP} collaboration, G.~Hinshaw et~al., \emph{{Nine-Year Wilkinson
  Microwave Anisotropy Probe (WMAP) Observations: Cosmological Parameter
  Results}},
  \href{http://dx.doi.org/10.1088/0067-0049/208/2/19}{\emph{Astrophys. J.
  Suppl.} {\bf 208} (2013) 19}, [\href{http://arxiv.org/abs/1212.5226}{{\tt
  1212.5226}}].

\bibitem{Planck:2018vyg}
{\scshape Planck} collaboration, N.~Aghanim et~al., \emph{{Planck 2018 results.
  VI. Cosmological parameters}},
  \href{http://dx.doi.org/10.1051/0004-6361/201833910}{\emph{Astron.
  Astrophys.} {\bf 641} (2020) A6},
  [\href{http://arxiv.org/abs/1807.06209}{{\tt 1807.06209}}].

\bibitem{Blumenthal:1984bp}
G.~R. Blumenthal, S.~M. Faber, J.~R. Primack and M.~J. Rees, \emph{{Formation
  of Galaxies and Large Scale Structure with Cold Dark Matter}},
  \href{http://dx.doi.org/10.1038/311517a0}{\emph{Nature} {\bf 311} (1984)
  517--525}.

\bibitem{Kolb:1990vq}
E.~W. Kolb and M.~S. Turner, \emph{{The Early Universe}}, vol.~69.
\newblock 1990,
  \href{http://dx.doi.org/10.1201/9780429492860}{10.1201/9780429492860}.

\bibitem{Jungman:1995df}
G.~Jungman, M.~Kamionkowski and K.~Griest, \emph{{Supersymmetric dark matter}},
  \href{http://dx.doi.org/10.1016/0370-1573(95)00058-5}{\emph{Phys. Rept.} {\bf
  267} (1996) 195--373}, [\href{http://arxiv.org/abs/hep-ph/9506380}{{\tt
  hep-ph/9506380}}].

\bibitem{Feng:2010gw}
J.~L. Feng, \emph{{Dark Matter Candidates from Particle Physics and Methods of
  Detection}},
  \href{http://dx.doi.org/10.1146/annurev-astro-082708-101659}{\emph{Ann. Rev.
  Astron. Astrophys.} {\bf 48} (2010) 495--545},
  [\href{http://arxiv.org/abs/1003.0904}{{\tt 1003.0904}}].

\bibitem{Arcadi:2017kky}
G.~Arcadi, M.~Dutra, P.~Ghosh, M.~Lindner, Y.~Mambrini, M.~Pierre et~al.,
  \emph{{The waning of the WIMP? A review of models, searches, and
  constraints}},
  \href{http://dx.doi.org/10.1140/epjc/s10052-018-5662-y}{\emph{Eur. Phys. J.
  C} {\bf 78} (2018) 203}, [\href{http://arxiv.org/abs/1703.07364}{{\tt
  1703.07364}}].

\bibitem{Roszkowski:2017nbc}
L.~Roszkowski, E.~M. Sessolo and S.~Trojanowski, \emph{{WIMP dark matter
  candidates and searches\textemdash{}current status and future prospects}},
  \href{http://dx.doi.org/10.1088/1361-6633/aab913}{\emph{Rept. Prog. Phys.}
  {\bf 81} (2018) 066201}, [\href{http://arxiv.org/abs/1707.06277}{{\tt
  1707.06277}}].

\bibitem{XENON:2018voc}
{\scshape XENON} collaboration, E.~Aprile et~al., \emph{{Dark Matter Search
  Results from a One Ton-Year Exposure of XENON1T}},
  \href{http://dx.doi.org/10.1103/PhysRevLett.121.111302}{\emph{Phys. Rev.
  Lett.} {\bf 121} (2018) 111302}, [\href{http://arxiv.org/abs/1805.12562}{{\tt
  1805.12562}}].

\bibitem{LUX:2017ree}
{\scshape LUX} collaboration, D.~S. Akerib et~al., \emph{{Limits on
  spin-dependent WIMP-nucleon cross section obtained from the complete LUX
  exposure}},
  \href{http://dx.doi.org/10.1103/PhysRevLett.118.251302}{\emph{Phys. Rev.
  Lett.} {\bf 118} (2017) 251302}, [\href{http://arxiv.org/abs/1705.03380}{{\tt
  1705.03380}}].

\bibitem{Abercrombie:2015wmb}
D.~Abercrombie et~al., \emph{{Dark Matter benchmark models for early LHC Run-2
  Searches: Report of the ATLAS/CMS Dark Matter Forum}},
  \href{http://dx.doi.org/10.1016/j.dark.2019.100371}{\emph{Phys. Dark Univ.}
  {\bf 27} (2020) 100371}, [\href{http://arxiv.org/abs/1507.00966}{{\tt
  1507.00966}}].

\bibitem{Bernal:2015xba}
N.~Bernal and X.~Chu, \emph{{$\mathbb {Z}_2$ SIMP Dark Matter}},
  \href{http://dx.doi.org/10.1088/1475-7516/2016/01/006}{\emph{JCAP} {\bf 1601}
  (2016) 006}, [\href{http://arxiv.org/abs/1510.08527}{{\tt 1510.08527}}].

\bibitem{Choi:2016hid}
S.-M. Choi and H.~M. Lee, \emph{{Resonant SIMP dark matter}},
  \href{http://dx.doi.org/10.1016/j.physletb.2016.04.055}{\emph{Phys. Lett.}
  {\bf B758} (2016) 47--53}, [\href{http://arxiv.org/abs/1601.03566}{{\tt
  1601.03566}}].

\bibitem{Choi:2016tkj}
S.-M. Choi, Y.-J. Kang and H.~M. Lee, \emph{{On thermal production of
  self-interacting dark matter}},
  \href{http://dx.doi.org/10.1007/JHEP12(2016)099}{\emph{JHEP} {\bf 12} (2016)
  099}, [\href{http://arxiv.org/abs/1610.04748}{{\tt 1610.04748}}].

\bibitem{Bernal:2017mqb}
N.~Bernal, X.~Chu and J.~Pradler, \emph{{Simply split strongly interacting
  massive particles}},
  \href{http://dx.doi.org/10.1103/PhysRevD.95.115023}{\emph{Phys. Rev.} {\bf
  D95} (2017) 115023}, [\href{http://arxiv.org/abs/1702.04906}{{\tt
  1702.04906}}].

\bibitem{Chauhan:2017eck}
B.~Chauhan, \emph{{Sub-MeV Self Interacting Dark Matter}},
  \href{http://dx.doi.org/10.1103/PhysRevD.97.123017}{\emph{Phys. Rev.} {\bf
  D97} (2018) 123017}, [\href{http://arxiv.org/abs/1711.02970}{{\tt
  1711.02970}}].

\bibitem{Mohanty:2019drv}
S.~Mohanty, A.~Patra and T.~Srivastava, \emph{{MeV scale model of SIMP dark
  matter, neutrino mass and leptogenesis}},
  \href{http://dx.doi.org/10.1088/1475-7516/2020/03/027}{\emph{JCAP} {\bf 03}
  (2020) 027}, [\href{http://arxiv.org/abs/1908.00909}{{\tt 1908.00909}}].

\bibitem{Bhattacharya:2019mmy}
S.~Bhattacharya, P.~Ghosh and S.~Verma, \emph{{SIMPler realisation of Scalar
  Dark Matter}},
  \href{http://dx.doi.org/10.1088/1475-7516/2020/01/040}{\emph{JCAP} {\bf 01}
  (2020) 040}, [\href{http://arxiv.org/abs/1904.07562}{{\tt 1904.07562}}].

\bibitem{Lee:2015gsa}
H.~M. Lee and M.-S. Seo, \emph{{Communication with SIMP dark mesons via Z'
  -portal}},
  \href{http://dx.doi.org/10.1016/j.physletb.2015.07.013}{\emph{Phys. Lett. B}
  {\bf 748} (2015) 316--322}, [\href{http://arxiv.org/abs/1504.00745}{{\tt
  1504.00745}}].

\bibitem{Yamanaka:2015tba}
N.~Yamanaka, S.~Fujibayashi, S.~Gongyo and H.~Iida, \emph{{Dark Matter in the
  Nonabelian Hidden Gauge Theory}},  in \emph{{2nd Toyama International
  Workshop on Higgs as a Probe of New Physics}}, 4, 2015.
\newblock \href{http://arxiv.org/abs/1504.08121}{{\tt 1504.08121}}.

\bibitem{Hochberg:2015vrg}
Y.~Hochberg, E.~Kuflik and H.~Murayama, \emph{{SIMP Spectroscopy}},
  \href{http://dx.doi.org/10.1007/JHEP05(2016)090}{\emph{JHEP} {\bf 05} (2016)
  090}, [\href{http://arxiv.org/abs/1512.07917}{{\tt 1512.07917}}].

\bibitem{Hochberg:2014kqa}
Y.~Hochberg, E.~Kuflik, H.~Murayama, T.~Volansky and J.~G. Wacker, \emph{{Model
  for Thermal Relic Dark Matter of Strongly Interacting Massive Particles}},
  \href{http://dx.doi.org/10.1103/PhysRevLett.115.021301}{\emph{Phys. Rev.
  Lett.} {\bf 115} (2015) 021301}, [\href{http://arxiv.org/abs/1411.3727}{{\tt
  1411.3727}}].

\bibitem{Choi:2017zww}
S.-M. Choi, Y.~Hochberg, E.~Kuflik, H.~M. Lee, Y.~Mambrini, H.~Murayama et~al.,
  \emph{{Vector SIMP dark matter}},
  \href{http://dx.doi.org/10.1007/JHEP10(2017)162}{\emph{JHEP} {\bf 10} (2017)
  162}, [\href{http://arxiv.org/abs/1707.01434}{{\tt 1707.01434}}].

\bibitem{Choi:2018jsb}
S.-M. Choi, H.~M. Lee, P.~Ko and A.~Natale, \emph{{Unitarizing SIMP scenario
  with dark vector resonances}},
  \href{http://dx.doi.org/10.22323/1.340.0036}{\emph{PoS} {\bf ICHEP2018}
  (2019) 036}, [\href{http://arxiv.org/abs/1811.02751}{{\tt 1811.02751}}].

\bibitem{Herms:2018ajr}
J.~Herms, A.~Ibarra and T.~Toma, \emph{{A new mechanism of sterile neutrino
  dark matter production}},
  \href{http://dx.doi.org/10.1088/1475-7516/2018/06/036}{\emph{JCAP} {\bf 06}
  (2018) 036}, [\href{http://arxiv.org/abs/1802.02973}{{\tt 1802.02973}}].

\bibitem{Ma:2015mjd}
E.~Ma, N.~Pollard, R.~Srivastava and M.~Zakeri, \emph{{Gauge $B-L$ Model with
  Residual $Z_3$ Symmetry}},
  \href{http://dx.doi.org/10.1016/j.physletb.2015.09.010}{\emph{Phys. Lett. B}
  {\bf 750} (2015) 135--138}, [\href{http://arxiv.org/abs/1507.03943}{{\tt
  1507.03943}}].

\bibitem{Thrane:2013oya}
E.~Thrane and J.~D. Romano, \emph{{Sensitivity curves for searches for
  gravitational-wave backgrounds}},
  \href{http://dx.doi.org/10.1103/PhysRevD.88.124032}{\emph{Phys. Rev. D} {\bf
  88} (2013) 124032}, [\href{http://arxiv.org/abs/1310.5300}{{\tt 1310.5300}}].

\bibitem{Sathyaprakash:2012jk}
B.~Sathyaprakash et~al., \emph{{Scientific Objectives of Einstein Telescope}},
  \href{http://dx.doi.org/10.1088/0264-9381/29/12/124013}{\emph{Class. Quant.
  Grav.} {\bf 29} (2012) 124013}, [\href{http://arxiv.org/abs/1206.0331}{{\tt
  1206.0331}}].

\bibitem{LISA:2017pwj}
{\scshape LISA} collaboration, P.~Amaro-Seoane et~al., \emph{{Laser
  Interferometer Space Antenna}},  \href{http://arxiv.org/abs/1702.00786}{{\tt
  1702.00786}}.

\bibitem{Crowder:2005nr}
J.~Crowder and N.~J. Cornish, \emph{{Beyond LISA: Exploring future
  gravitational wave missions}},
  \href{http://dx.doi.org/10.1103/PhysRevD.72.083005}{\emph{Phys. Rev. D} {\bf
  72} (2005) 083005}, [\href{http://arxiv.org/abs/gr-qc/0506015}{{\tt
  gr-qc/0506015}}].

\bibitem{Seto:2001qf}
N.~Seto, S.~Kawamura and T.~Nakamura, \emph{{Possibility of direct measurement
  of the acceleration of the universe using 0.1-Hz band laser interferometer
  gravitational wave antenna in space}},
  \href{http://dx.doi.org/10.1103/PhysRevLett.87.221103}{\emph{Phys. Rev.
  Lett.} {\bf 87} (2001) 221103},
  [\href{http://arxiv.org/abs/astro-ph/0108011}{{\tt astro-ph/0108011}}].

\bibitem{Sato:2017dkf}
S.~Sato et~al., \emph{{The status of DECIGO}},
  \href{http://dx.doi.org/10.1088/1742-6596/840/1/012010}{\emph{J. Phys. Conf.
  Ser.} {\bf 840} (2017) 012010}.

\bibitem{Lentati:2015qwp}
L.~Lentati et~al., \emph{{European Pulsar Timing Array Limits On An Isotropic
  Stochastic Gravitational-Wave Background}},
  \href{http://dx.doi.org/10.1093/mnras/stv1538}{\emph{Mon. Not. Roy. Astron.
  Soc.} {\bf 453} (2015) 2576--2598},
  [\href{http://arxiv.org/abs/1504.03692}{{\tt 1504.03692}}].

\bibitem{Janssen:2014dka}
G.~Janssen et~al., \emph{{Gravitational wave astronomy with the SKA}},
  \href{http://dx.doi.org/10.22323/1.215.0037}{\emph{PoS} {\bf AASKA14} (2015)
  037}, [\href{http://arxiv.org/abs/1501.00127}{{\tt 1501.00127}}].

\bibitem{Fairbairn:2019xog}
M.~Fairbairn, E.~Hardy and A.~Wickens, \emph{{Hearing without seeing:
  gravitational waves from hot and cold hidden sectors}},
  \href{http://dx.doi.org/10.1007/JHEP07(2019)044}{\emph{JHEP} {\bf 07} (2019)
  044}, [\href{http://arxiv.org/abs/1901.11038}{{\tt 1901.11038}}].

\bibitem{Breitbach:2018ddu}
M.~Breitbach, J.~Kopp, E.~Madge, T.~Opferkuch and P.~Schwaller, \emph{{Dark,
  Cold, and Noisy: Constraining Secluded Hidden Sectors with Gravitational
  Waves}}, \href{http://dx.doi.org/10.1088/1475-7516/2019/07/007}{\emph{JCAP}
  {\bf 07} (2019) 007}, [\href{http://arxiv.org/abs/1811.11175}{{\tt
  1811.11175}}].

\bibitem{Schwaller:2015tja}
P.~Schwaller, \emph{{Gravitational Waves from a Dark Phase Transition}},
  \href{http://dx.doi.org/10.1103/PhysRevLett.115.181101}{\emph{Phys. Rev.
  Lett.} {\bf 115} (2015) 181101}, [\href{http://arxiv.org/abs/1504.07263}{{\tt
  1504.07263}}].

\bibitem{Jaeckel:2016jlh}
J.~Jaeckel, V.~V. Khoze and M.~Spannowsky, \emph{{Hearing the signal of dark
  sectors with gravitational wave detectors}},
  \href{http://dx.doi.org/10.1103/PhysRevD.94.103519}{\emph{Phys. Rev. D} {\bf
  94} (2016) 103519}, [\href{http://arxiv.org/abs/1602.03901}{{\tt
  1602.03901}}].

\bibitem{Addazi:2016fbj}
A.~Addazi, \emph{{Limiting First Order Phase Transitions in Dark Gauge Sectors
  from Gravitational Waves experiments}},
  \href{http://dx.doi.org/10.1142/S0217732317500493}{\emph{Mod. Phys. Lett. A}
  {\bf 32} (2017) 1750049}, [\href{http://arxiv.org/abs/1607.08057}{{\tt
  1607.08057}}].

\bibitem{Baldes:2017rcu}
I.~Baldes, \emph{{Gravitational waves from the asymmetric-dark-matter
  generating phase transition}},
  \href{http://dx.doi.org/10.1088/1475-7516/2017/05/028}{\emph{JCAP} {\bf 05}
  (2017) 028}, [\href{http://arxiv.org/abs/1702.02117}{{\tt 1702.02117}}].

\bibitem{Tsumura:2017knk}
K.~Tsumura, M.~Yamada and Y.~Yamaguchi, \emph{{Gravitational wave from dark
  sector with dark pion}},
  \href{http://dx.doi.org/10.1088/1475-7516/2017/07/044}{\emph{JCAP} {\bf 07}
  (2017) 044}, [\href{http://arxiv.org/abs/1704.00219}{{\tt 1704.00219}}].

\bibitem{Baldes:2018emh}
I.~Baldes and C.~Garcia-Cely, \emph{{Strong gravitational radiation from a
  simple dark matter model}},
  \href{http://dx.doi.org/10.1007/JHEP05(2019)190}{\emph{JHEP} {\bf 05} (2019)
  190}, [\href{http://arxiv.org/abs/1809.01198}{{\tt 1809.01198}}].

\bibitem{Croon:2018erz}
D.~Croon, V.~Sanz and G.~White, \emph{{Model Discrimination in Gravitational
  Wave spectra from Dark Phase Transitions}},
  \href{http://dx.doi.org/10.1007/JHEP08(2018)203}{\emph{JHEP} {\bf 08} (2018)
  203}, [\href{http://arxiv.org/abs/1806.02332}{{\tt 1806.02332}}].

\bibitem{Kamionkowski:1993fg}
M.~Kamionkowski, A.~Kosowsky and M.~S. Turner, \emph{{Gravitational radiation
  from first order phase transitions}},
  \href{http://dx.doi.org/10.1103/PhysRevD.49.2837}{\emph{Phys. Rev. D} {\bf
  49} (1994) 2837--2851}, [\href{http://arxiv.org/abs/astro-ph/9310044}{{\tt
  astro-ph/9310044}}].

\bibitem{Witten:1984rs}
E.~Witten, \emph{{Cosmic Separation of Phases}},
  \href{http://dx.doi.org/10.1103/PhysRevD.30.272}{\emph{Phys. Rev. D} {\bf 30}
  (1984) 272--285}.

\bibitem{Hogan:1986qda}
C.~J. Hogan, \emph{{Gravitational radiation from cosmological phase
  transitions}}, {\emph{Mon. Not. Roy. Astron. Soc.} {\bf 218} (1986)
  629--636}.

\bibitem{Grojean:2006bp}
C.~Grojean and G.~Servant, \emph{{Gravitational Waves from Phase Transitions at
  the Electroweak Scale and Beyond}},
  \href{http://dx.doi.org/10.1103/PhysRevD.75.043507}{\emph{Phys. Rev. D} {\bf
  75} (2007) 043507}, [\href{http://arxiv.org/abs/hep-ph/0607107}{{\tt
  hep-ph/0607107}}].

\bibitem{Pierre:2018man}
M.~M. Pierre, \emph{{Dark Matter phenomenology : from simplified WIMP models to
  refined alternative solutions}}.
\newblock PhD thesis, U. Paris-Saclay, 2018.
\newblock \href{http://arxiv.org/abs/1901.05822}{{\tt 1901.05822}}.

\bibitem{Bhattacharya:2016ysw}
S.~Bhattacharya, P.~Poulose and P.~Ghosh, \emph{{Multipartite Interacting
  Scalar Dark Matter in the light of updated LUX data}},
  \href{http://dx.doi.org/10.1088/1475-7516/2017/04/043}{\emph{JCAP} {\bf 04}
  (2017) 043}, [\href{http://arxiv.org/abs/1607.08461}{{\tt 1607.08461}}].

\bibitem{Weinberg}
S.~Weinberg, \emph{Gauge and global symmetries at high temperature},
  \href{http://dx.doi.org/10.1103/PhysRevD.9.3357}{\emph{Phys. Rev. D} {\bf 9}
  (Jun, 1974) 3357--3378}.

\bibitem{Dolan}
L.~Dolan and R.~Jackiw, \emph{Symmetry behavior at finite temperature},
  \href{http://dx.doi.org/10.1103/PhysRevD.9.3320}{\emph{Phys. Rev. D} {\bf 9}
  (Jun, 1974) 3320--3341}.

\bibitem{Kirzhnits:1976ts}
D.~A. Kirzhnits and A.~D. Linde, \emph{{Symmetry Behavior in Gauge Theories}},
  \href{http://dx.doi.org/10.1016/0003-4916(76)90279-7}{\emph{Annals Phys.}
  {\bf 101} (1976) 195--238}.

\bibitem{Carrington:1991hz}
M.~E. Carrington, \emph{{The Effective potential at finite temperature in the
  Standard Model}},
  \href{http://dx.doi.org/10.1103/PhysRevD.45.2933}{\emph{Phys. Rev. D} {\bf
  45} (1992) 2933--2944}.

\bibitem{Gross}
D.~J. Gross, R.~D. Pisarski and L.~G. Yaffe, \emph{Qcd and instantons at finite
  temperature}, \href{http://dx.doi.org/10.1103/RevModPhys.53.43}{\emph{Rev.
  Mod. Phys.} {\bf 53} (Jan, 1981) 43--80}.

\bibitem{Fendley:1987ef}
P.~Fendley, \emph{{The Effective Potential and the Coupling Constant at High
  Temperature}},
  \href{http://dx.doi.org/10.1016/0370-2693(87)90599-5}{\emph{Phys. Lett. B}
  {\bf 196} (1987) 175--180}.

\bibitem{Kapusta}
J.~Kapusta, \emph{{Finite temperature Field Theory}}, {\emph{Cambridge
  University Press} (1989) }.

\bibitem{Arnold:1992rz}
P.~B. Arnold and O.~Espinosa, \emph{{The Effective potential and first order
  phase transitions: Beyond leading-order}},
  \href{http://dx.doi.org/10.1103/PhysRevD.47.3546}{\emph{Phys. Rev. D} {\bf
  47} (1993) 3546}, [\href{http://arxiv.org/abs/hep-ph/9212235}{{\tt
  hep-ph/9212235}}].

\bibitem{Parwani:1991gq}
R.~R. Parwani, \emph{{Resummation in a hot scalar field theory}},
  \href{http://dx.doi.org/10.1103/PhysRevD.45.4695}{\emph{Phys. Rev. D} {\bf
  45} (1992) 4695}, [\href{http://arxiv.org/abs/hep-ph/9204216}{{\tt
  hep-ph/9204216}}].

\bibitem{Quiros:1999jp}
M.~Quiros, \emph{{Finite temperature field theory and phase transitions}},  in
  \emph{{ICTP Summer School in High-Energy Physics and Cosmology}},
  pp.~187--259, 1, 1999.
\newblock \href{http://arxiv.org/abs/hep-ph/9901312}{{\tt hep-ph/9901312}}.

\bibitem{Athron:2020sbe}
P.~Athron, C.~Bal\'azs, A.~Fowlie and Y.~Zhang, \emph{{PhaseTracer: tracing
  cosmological phases and calculating transition properties}},
  \href{http://dx.doi.org/10.1140/epjc/s10052-020-8035-2}{\emph{Eur. Phys. J.
  C} {\bf 80} (2020) 567}, [\href{http://arxiv.org/abs/2003.02859}{{\tt
  2003.02859}}].

\bibitem{Kosowsky}
A.~Kosowsky, M.~S. Turner and R.~Watkins, \emph{Gravitational radiation from
  colliding vacuum bubbles},
  \href{http://dx.doi.org/10.1103/PhysRevD.45.4514}{\emph{Phys. Rev. D} {\bf
  45} (Jun, 1992) 4514--4535}.

\bibitem{Turner}
A.~Kosowsky and M.~S. Turner, \emph{Gravitational radiation from colliding
  vacuum bubbles: Envelope approximation to many-bubble collisions},
  \href{http://dx.doi.org/10.1103/PhysRevD.47.4372}{\emph{Phys. Rev. D} {\bf
  47} (May, 1993) 4372--4391}.

\bibitem{Huber_2008}
S.~J. Huber and T.~Konstandin, \emph{Gravitational wave production by
  collisions: more bubbles},
  \href{http://dx.doi.org/10.1088/1475-7516/2008/09/022}{\emph{Journal of
  Cosmology and Astroparticle Physics} {\bf 2008} (Sep, 2008) 022}.

\bibitem{Watkins}
A.~Kosowsky, M.~S. Turner and R.~Watkins, \emph{Gravitational waves from
  first-order cosmological phase transitions},
  \href{http://dx.doi.org/10.1103/PhysRevLett.69.2026}{\emph{Phys. Rev. Lett.}
  {\bf 69} (Oct, 1992) 2026--2029}.

\bibitem{Marc}
M.~Kamionkowski, A.~Kosowsky and M.~S. Turner, \emph{Gravitational radiation
  from first-order phase transitions},
  \href{http://dx.doi.org/10.1103/PhysRevD.49.2837}{\emph{Phys. Rev. D} {\bf
  49} (Mar, 1994) 2837--2851}.

\bibitem{Caprini_2008}
C.~Caprini, R.~Durrer and G.~Servant, \emph{Gravitational wave generation from
  bubble collisions in first-order phase transitions: An analytic approach},
  \href{http://dx.doi.org/10.1103/physrevd.77.124015}{\emph{Physical Review D}
  {\bf 77} (Jun, 2008) }.

\bibitem{Hindmarsh}
M.~Hindmarsh, S.~J. Huber, K.~Rummukainen and D.~J. Weir, \emph{Gravitational
  waves from the sound of a first order phase transition},
  \href{http://dx.doi.org/10.1103/PhysRevLett.112.041301}{\emph{Phys. Rev.
  Lett.} {\bf 112} (Jan, 2014) 041301}.

\bibitem{Leitao:2012tx}
L.~Leitao, A.~Megevand and A.~D. Sanchez, \emph{{Gravitational waves from the
  electroweak phase transition}},
  \href{http://dx.doi.org/10.1088/1475-7516/2012/10/024}{\emph{JCAP} {\bf 10}
  (2012) 024}, [\href{http://arxiv.org/abs/1205.3070}{{\tt 1205.3070}}].

\bibitem{Giblin:2013kea}
J.~T. Giblin, Jr. and J.~B. Mertens, \emph{{Vacuum Bubbles in the Presence of a
  Relativistic Fluid}},
  \href{http://dx.doi.org/10.1007/JHEP12(2013)042}{\emph{JHEP} {\bf 12} (2013)
  042}, [\href{http://arxiv.org/abs/1310.2948}{{\tt 1310.2948}}].

\bibitem{Giblin:2014qia}
J.~T. Giblin and J.~B. Mertens, \emph{{Gravitional radiation from first-order
  phase transitions in the presence of a fluid}},
  \href{http://dx.doi.org/10.1103/PhysRevD.90.023532}{\emph{Phys. Rev. D} {\bf
  90} (2014) 023532}, [\href{http://arxiv.org/abs/1405.4005}{{\tt 1405.4005}}].

\bibitem{Hindmarsh_2015}
M.~Hindmarsh, S.~J. Huber, K.~Rummukainen and D.~J. Weir, \emph{Numerical
  simulations of acoustically generated gravitational waves at a first order
  phase transition},
  \href{http://dx.doi.org/10.1103/physrevd.92.123009}{\emph{Physical Review D}
  {\bf 92} (Dec, 2015) }.

\bibitem{Chiara}
C.~Caprini and R.~Durrer, \emph{Gravitational waves from stochastic
  relativistic sources: Primordial turbulence and magnetic fields},
  \href{http://dx.doi.org/10.1103/PhysRevD.74.063521}{\emph{Phys. Rev. D} {\bf
  74} (Sep, 2006) 063521}.

\bibitem{Kahniashvili}
T.~Kahniashvili, A.~Kosowsky, G.~Gogoberidze and Y.~Maravin,
  \emph{Detectability of gravitational waves from phase transitions},
  \href{http://dx.doi.org/10.1103/PhysRevD.78.043003}{\emph{Phys. Rev. D} {\bf
  78} (Aug, 2008) 043003}.

\bibitem{Kahniashvili:2008pe}
T.~Kahniashvili, L.~Campanelli, G.~Gogoberidze, Y.~Maravin and B.~Ratra,
  \emph{{Gravitational Radiation from Primordial Helical Inverse Cascade MHD
  Turbulence}}, \href{http://dx.doi.org/10.1103/PhysRevD.78.123006}{\emph{Phys.
  Rev. D} {\bf 78} (2008) 123006}, [\href{http://arxiv.org/abs/0809.1899}{{\tt
  0809.1899}}].

\bibitem{Kahniashvili:2009mf}
T.~Kahniashvili, L.~Kisslinger and T.~Stevens, \emph{{Gravitational Radiation
  Generated by Magnetic Fields in Cosmological Phase Transitions}},
  \href{http://dx.doi.org/10.1103/PhysRevD.81.023004}{\emph{Phys. Rev. D} {\bf
  81} (2010) 023004}, [\href{http://arxiv.org/abs/0905.0643}{{\tt 0905.0643}}].

\bibitem{Caprini:2009yp}
C.~Caprini, R.~Durrer and G.~Servant, \emph{{The stochastic gravitational wave
  background from turbulence and magnetic fields generated by a first-order
  phase transition}},
  \href{http://dx.doi.org/10.1088/1475-7516/2009/12/024}{\emph{JCAP} {\bf 12}
  (2009) 024}, [\href{http://arxiv.org/abs/0909.0622}{{\tt 0909.0622}}].

\bibitem{Kudoh:2005as}
H.~Kudoh, A.~Taruya, T.~Hiramatsu and Y.~Himemoto, \emph{{Detecting a
  gravitational-wave background with next-generation space interferometers}},
  \href{http://dx.doi.org/10.1103/PhysRevD.73.064006}{\emph{Phys. Rev. D} {\bf
  73} (2006) 064006}, [\href{http://arxiv.org/abs/gr-qc/0511145}{{\tt
  gr-qc/0511145}}].

\end{thebibliography}\endgroup

\end{document}